\documentclass[english,aps,prx,superscriptaddress,twocolumn,address,showacknowledgments]{revtex4}
\usepackage[T1]{fontenc}
\usepackage[latin9]{inputenc}
\usepackage{babel}
\usepackage{amsmath}
\usepackage{amssymb}
\usepackage{graphicx}
\usepackage{esint}
\PassOptionsToPackage{normalem}{ulem}
\usepackage{ulem}
\usepackage[unicode=true,pdfusetitle,
 bookmarks=true,bookmarksnumbered=false,bookmarksopen=false,
 breaklinks=false,pdfborder={0 0 1},backref=false,colorlinks=false]
 {hyperref}

\makeatletter
\@ifundefined{textcolor}{}
{%
 \definecolor{BLACK}{gray}{0}
 \definecolor{WHITE}{gray}{1}
 \definecolor{RED}{rgb}{1,0,0}
 \definecolor{GREEN}{rgb}{0,1,0}
 \definecolor{BLUE}{rgb}{0,0,1}
 \definecolor{CYAN}{cmyk}{1,0,0,0}
 \definecolor{MAGENTA}{cmyk}{0,1,0,0}
 \definecolor{YELLOW}{cmyk}{0,0,1,0}
}

\makeatother

\begin{document}

\title{Optomechanical creation of magnetic fields for photons on a lattice}

\author{M. Schmidt}

\address{University of Erlangen-N\"urnberg, Staudtstr. 7, Institute for Theoretical
Physics, D-91058 Erlangen, Germany}

\author{S. Ke{\ss}ler}

\address{University of Erlangen-N\"urnberg, Staudtstr. 7, Institute for Theoretical
Physics, D-91058 Erlangen, Germany}

\author{V. Peano}

\address{University of Erlangen-N\"urnberg, Staudtstr. 7, Institute for Theoretical
Physics, D-91058 Erlangen, Germany}

\author{O. Painter}

\address{Institute for Quantum Information and Matter and Thomas J. Watson,
Sr., Laboratory of Applied Physics, California Institute of Technology,
Pasadena, CA 91125, USA}

\author{F. Marquardt}

\address{University of Erlangen-N\"urnberg, Staudtstr. 7, Institute for Theoretical
Physics, D-91058 Erlangen, Germany}

\address{Max Planck Institute for the Science of Light, G\"unther-Scharowsky-Stra\ss e
1/Bau 24, D-91058 Erlangen, Germany}
\begin{abstract}
We propose using the optomechanical interaction to create artificial
magnetic fields for photons on a lattice. The ingredients required
are an optomechanical crystal, i.e. a piece of dielectric with the
right pattern of holes, and two laser beams with the right pattern
of phases. One of the two proposed schemes is based on optomechanical
modulation of the links between optical modes, while the other is
a lattice extension of optomechanical wavelength-conversion setups.
We illustrate the resulting optical spectrum, photon transport in
the presence of an artificial Lorentz force, edge states, and the
photonic Aharonov-Bohm effect. Moreover, we briefly describe the gauge
fields acting on the synthetic dimension related to the phonon/photon
degree of freedom. 
\end{abstract}
\maketitle
Light interacting with nano-mechanical motion via the radiation pressure
force is studied in the field of optomechanics. The field has seen
rapid progress in the last few years (see the recent review \cite{Aspelmeyer2013RMPArxiv}).
So far, most experimental achievements have been realized in setups
comprising one optical mode coupled to one vibrational mode. Obviously,
one of the next frontiers will be the combination of many such optomechanical
cells into an optomechanical array, enabling the optical in-situ investigation
of (quantum) many-body dynamics of interacting photons and phonons.
Many experimental platforms allow to be scaled up to arrays. However,
optomechanical crystals seem to be the best suited candidate at the
present stage. Optomechanical crystals are formed by the periodic
spatial patterning of regular dielectric and elastic materials, resulting
in an enhanced coupling between optical and acoustic waves via moving
boundary or electrostriction radiation pressure effects. Two-dimensional
(2D) optomechanical crystals with both photonic and phononic bandgaps
\cite{Safavi-Naeini2010} can be fabricated by standard microfabrication
techniques through the lithographic patterning, plasma etching, and
release of a thin-film material \cite{Eichenfield2009}. These 2D
crystals for light and sound can be used to create a circuit architecture
for the routing and localization of photons and phonons \cite{Eichenfield2009,Safavi-Naeini2010APL,Gavartin2011PRL_OMC,Chan2011Cooling,SafaviNaeini2014SnowCavity}.

Optomechanical arrays promise to be a versatile platform for exploring
optomechanical many-body physics. Several aspects have already been
investigated theoretically, e.g. synchronization \cite{Heinrich2011CollDyn,Holmes2012,Ludwig2013},
long-range interactions \cite{Bhattacharya2008,Xuereb2012}, reservoir
engineering \cite{Tombadin2012}, entanglement \cite{Schmidt2012,Akram2012},
correlated quantum many-body states \cite{Ludwig2013}, slow light
\cite{Chang2011}, transport in a 1D chain \cite{Chen2014}, and graphene-like
Dirac physics \cite{Schmidt2014}. . 

One of the central aims in photonics is to build waveguides that are
robust against disorder and do not display backscattering. Recently
there have been several proposals \cite{Koch2011,Hafezi2011,Umucallar2011,Fang2012,Hafezi2012OptExpr}\textbf{
}to engineer non-reciprocal transport for photons. On the lattice,
this corresponds to an artificial magnetic field,which would (among
other effects) enable chiral edge states that display the desired
robustness against disorder. First experiments have shown such edge
states \cite{Hafezi2013,Mittal2014,Rechtsman2013Nat}. These developments
in photonics are related to a growing effort across various fields
to produce synthetic gauge fields for neutral particles \cite{Bermudez2011,Aidelsburger2013,Miyake2013}.

In this paper we will propose two schemes to generate an artificial
magnetic field for photons on a lattice. In contrast to any previous
proposals or experiments for photonic magnetic fields on a lattice,
these would be controlled all-optically and, crucially, they would
be tunable in-situ by changing the properties of a laser field (frequency,
intensity, and phase pattern). They require no more than a patterned
dielectric slab illuminated by two laser beams with suitably engineered
optical phase fields. The crucial ingredient is the optomechanical
interaction.

On the classical level, a charged particle subject to a magnetic field
experiences a Lorentz force. In the quantum regime, the appearance
of Landau levels leads to the integer and fractional quantum Hall
effects, where topologically protected chiral edge states are responsible
for a quantized Hall conductance. On a closed orbit, a particle with
charge $q$ will pick up a phase that is given by the magnetic flux
$\Phi$ through the circumscribed area, where $\Phi=(q/\hbar)\int{\bf B}\cdot d\mathbf{S}$ in
units of the flux quantum, with ${\bf B}$ denoting the magnetic field. On
a lattice, a charged particle hopping from site i to j acquires a
Peierls phase $\phi_{ij}=(q/\hbar)\int_{{\bf r}_{i}}^{{\bf r}_{j}}{\bf A}d{\bf r}$
determined by the vector potential ${\bf A}$. Conversely, if we can
engineer a Hamiltonian for neutral particles containing arbitrary
Peierls phases, 
\begin{equation}
\hat{H}_{\text{hop}}=\hbar J\sum_{\langle ij\rangle}e^{i\phi_{ij}}\hat{a}_{j}^{\dagger}\hat{a}_{i}+\text{h.c.},\label{eq:HopEffective}
\end{equation}
we are able to produce a synthetic magnetic field. Here $\hat{a}_{i}$
is the (bosonic) annihilation operator on lattice site i. We note
in passing that different phase configurations can lead to identical
flux patterns, reflecting the gauge invariance of Maxwell's equations
under the transformation ${\bf A}\rightarrow{\bf A}+\nabla\xi({\bf r})$
for any scalar function $\xi$.

Every defect in an optomechanical crystal \cite{Eichenfield2009,Safavi-Naeini2010APL,Gavartin2011PRL_OMC,Chan2011Cooling,SafaviNaeini2014SnowCavity}
supports a localized vibrational (annihilation operator $\hat{b}$,
eigenfrequency $\Omega_{0}$) and optical mode ($\hat{a}$, frequency
$\omega_{\text{cav}}$) that interact via radiation pressure, giving
rise to the standard optomechanical interaction \cite{Aspelmeyer2013RMPArxiv}:
\begin{equation}
\hat{H}_{\text{int}}=-\hbar g_{0}\hat{a}^{\dagger}\hat{a}(\hat{b}^{\dagger}+\hat{b}).\label{eq:OMInt}
\end{equation}
This can be utilized in two basic ways to introduce phases for the
hopping of photons. First, one can drive the optical mode by a control
laser (frequency $\omega_{L}$) close to the red sideband, $\omega_{L}\sim\omega_{\text{cav}}-\Omega_{0}$.
Following the standard procedure of linearization and rotating wave
approximation (RWA) \cite{Aspelmeyer2013RMPArxiv} one recovers a
swap Hamiltonian, $g\hat{a}^{\dagger}\hat{b}+\text{h.c.}$, in which
the phase $\phi$ of the coupling $g\in\mathbb{C}$ is set by the
control laser phase. We will show below how this can be used to create
a photonic gauge field. There is, however, also a second route, namely,
driving the vibrational mode into a large amplitude coherent state,
$\langle\hat{b}(t)\rangle=|\beta|e^{-i(\Omega t+\phi)}$, using the
radiation pressure force. These oscillations then weakly modulate
the optical eigenfrequency, $\omega_{\text{cav}}(t)=\omega_{\text{cav},0}+2g_{0}|\beta|\cos(\Omega t+\phi)$,
with the phase $\phi$ set by the oscillations. Again, in a suitable
setting this will lead to an artificial magnetic field for the photons.
We now describe both methods in turn. 

\begin{figure}
\includegraphics[width=1\columnwidth]{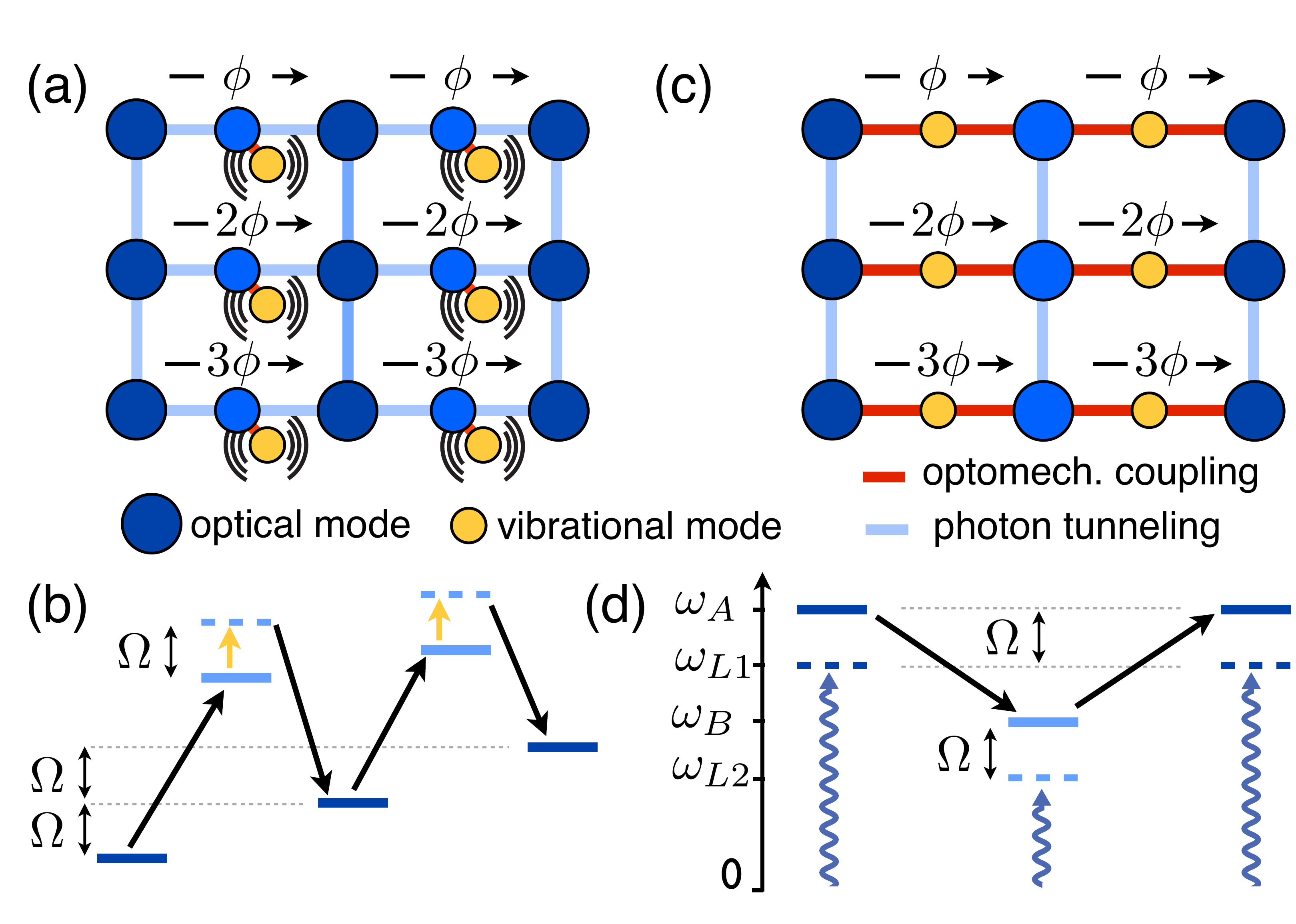}

\protect\caption{Proposed schemes to create a photonic gauge field in optomechanical
arrays by engineering photon hopping phases. (a) Modulated link scheme.
(b) Corresponding optical spectrum of a row with relevant sidebands
(dashed). Driven vibrational modes (yellow) optomechanically modulate
the frequency of optical link modes. Tunneling photons are thus up-converted
to the first sideband and pick up the phase of the modulation. Arrows
in (b) indicate the resonant photon transmission process in a row.
(c) Wavelength conversion scheme and (d) corresponding optical spectrum:
Neighboring modes in a row couple to a vibrational mode (yellow) optomechanically
(red lines, denoting the linearized optomechanical interaction). Two
lasers, driving the optical modes close to the red sidebands (wiggly
arrows in d), give rise to resonant photon-phonon-photon conversion
with the phase set by the lasers. Different rows are connected via
simple photon hopping (blue lines), without a phase, in both schemes
(a,c). The indicated phase configuration corresponds to a constant
magnetic field. \label{fig:Schemes}}
\end{figure}

\begin{figure*}
\includegraphics[width=1.95\columnwidth]{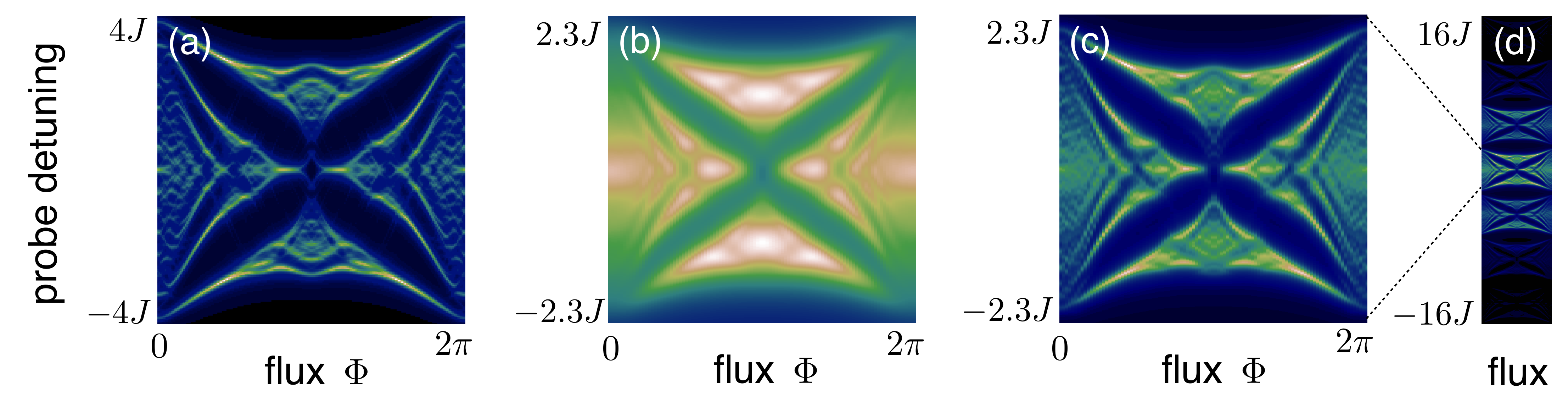}

\protect\caption{Comparison between the experimentally accessible optical spectrum
(LDOS) of the ideal effective Hofstadter model (a) and the actual
results from the proposed modulated link scheme (b-d), for different
optical decay rates and magnetic fluxes. The simulation results indicate
that the scheme works even beyond the perturbative regime. The resulting
Hofstadter butterfly could be observed by a local tapered fiber probe.
Modulation of links produces higher side bands (d). The phase configuration
corresponds to a constant magnetic flux $\Phi$ per plaquette, see
Fig. \ref{fig:Schemes}. {[}Parameters: grid: 10x10 (a), 12x12 (b-d);
$J_{\text{eff}}=0.108\Omega_{0}$ (a-d); $\kappa=0.01\Omega_{0}$
(a,c,d), $\kappa=0.05\Omega_{0}$ (b); $J=0.3\Omega_{0}$ (b-d); $g_{0}|\beta|=0.3\Omega_{0}$
(b-d); optical eigenfrequencies (relative to first mode) in a row
(left to right) including interface modes: $(0,0.5,1,1.5,\dots)\Omega_{0}$
(b-d).{]} \label{fig:OptomechanicalButterfly}}
\end{figure*}

\emph{Modulated link scheme}. \textendash{} Recently Fang et. al.
\cite{Fang2012,Tzuang2014} proposed to create a photonic gauge field
by electro-optically modulating the photon hopping rate $J_{ij}=J\cos(\Omega t+\phi_{ij}$)
between neighboring cavities. This would require locally wired electrodes
for each link of the lattice. Here we propose a potentially more powerful
all-optical implementation of that idea. We employ optomechanically
driven photon transitions, as first discussed in \cite{Heinrich2010_Landau},
but extended to a scheme with modulated interface modes, depicted
in Figure \ref{fig:Schemes} (a). We now discuss the leftmost three
optical modes in the first row, $\hat{a}_{A}$, $\hat{a}_{I}$, $\hat{a}_{B}$
(from left to right), exemplary of the full grid. Their coherent dynamics
is governed by the Hamiltonian 
\begin{equation}
\hat{H}/\hbar=\sum_{i=A,B}\omega_{i}\hat{a}_{i}^{\dagger}\hat{a}_{i}+\omega_{I}(t)\hat{a}_{I}^{\dagger}\hat{a}_{I}-J(\hat{a}_{I}^{\dagger}\hat{a}_{A}+\hat{a}_{B}^{\dagger}\hat{a}_{I}+\text{h.c.}).\label{eq:FloquetSchemeTimeDepH}
\end{equation}
The terms describe, in this order, the first (A) and third (B) optical
mode, the temporally modulated interface mode (I), and its tight binding
coupling to the neighboring A and B modes with photon tunneling rate
$J$. As discussed further below, the eigenfrequency $\bar{\omega}_{I}$
of the interface mode should be well separated from the eigenfrequencies
of the adjacent A and B modes, for the transition $A\text{\textendash I\textendash B}$
to be virtual. The interface mode is optomechanically coupled to a
mechanical mode, which itself is driven into a large amplitude coherent
state. As mentioned above, this gives rise to a weak modulation of
its optical eigenfrequency, $\omega_{I}(t)=\bar{\omega}_{I}+2g_{0}|\beta|\cos(\Omega t+\phi)$,
with the phase $\phi$ set by the driving.  The required mechanical
driving is easily generated by two-tone laser excitation at a frequency
difference $\Omega$. The beating between the laser beams gives rise
to a sinusoidal radiation pressure force, which drives the mechanical
mode. If $\omega_{B}=\omega_{A}+\Omega$, then a photon hopping from
site A to B picks up the phase $\phi$ of the modulation: Starting
from $\hat{a}_{A}$, it tunnels into $\hat{a}_{I}$ where it is inelastically
up-scattered into the first sideband by the modulation and subsequently
tunnels into $\hat{a}_{B}$ resonantly, as shown by the spectrum in
Figure \ref{fig:Schemes}(b). We can derive an effective Hamiltonian,
$\hbar J_{{\rm eff}}e^{i\phi}\hat{a}{}_{B}^{\dagger}\hat{a}{}_{A}+\text{h.c.}$
for this process by integrating out the interface mode $I$ using
Floquet perturbative methods to third order (see Appendix \ref{sec:detailsFloquetHamiltonian}). For the effective hopping rate we find $J_{{\rm eff}}=g_{0}|\beta|J^{2}/[(\omega_{A}-\omega_{I})(\omega_{B}-\omega_{I})]$,
to leading order in $J$ and $\left|\beta\right|$. Concatenating
such three-mode blocks, we create a linear chain (the first row in
Figure \ref{fig:Schemes}a), with its optical spectrum schematically
depicted in Figure \ref{fig:Schemes}(b). Every time a photon hops
to the right, it is up-converted and picks up the phase of the drive.
To obtain a 2d grid, we stack identical chains and connect neighboring
rows by direct photon hopping (whose rate must be chosen to equal
$J_{\text{eff}}$, to obtain isotropic hopping), as depicted in Figure
\ref{fig:Schemes}(b). The phase configuration in Figure \ref{fig:Schemes}(a)
corresponds to a constant magnetic field. Note that in contrast to
the general Hamiltonian (\ref{eq:HopEffective}), this scheme does
not allow for phases when hopping between rows, yet it is still possible
to achieve an arbitrary flux through every plaquette. Hence, arbitrary
magnetic fields can be generated, provided one can control the driving
laser phase at every interface mode. With the help of wave front engineering,
this can be achieved with no more than two lasers: A homogeneous 'carrier'
beam $E_{1}=E_{10}e^{-i\omega_{L}t}$ and a 'modulation' beam $E_{2}=E_{20}e^{-i(\omega_{L}+\Omega)t-i\phi(x,y)}$,
with an imprinted phase pattern $\phi(x,y)$. Interference yields
the desired temporally modulated intensity $\left|E_{10}\right|^{2}+\left|E_{20}\right|^{2}+2{\rm Re}[E_{10}^{*}E_{20}e^{-i(\omega_{{\rm L}}t+\phi(x,y))}]$,
exerting a radiation force with a site-dependent phase. Care has to
be taken to avoid exciting other vibrational modes (those not at the
interface mode), by engineering them to have different mechanical
frequencies. To this end, the driving frequency $\Omega$ would usually
be chosen close to the mechanical eigenfrequency $\Omega_{0}$, so
the mechanical amplitude is enhanced by the mechanical quality factor
and is thus much larger than any spurious amplitude in other (off-resonant)
modes. By engineering the intensity pattern $|E_{20}(x,y)|$ as well,
one could suppress any such unwanted effects even further.

\emph{Wavelength conversion scheme}. - There is another, alternative
way of engineering an optical Peierls phase, and it is related to
optomechanical wavelength conversion \cite{Hill2012WavelengthConversion,Dong2012}.
In wavelength conversion setups, low frequency photons in one mode
are up-converted to a higher frequency in another mode by exploiting
the modes' mutual optomechanical coupling to a vibrational mode. We
propose to scale up this idea into a grid as depicted in Figure \ref{fig:Schemes}(c).
The leftmost three modes in the first row depict (in this order) an
optical mode (annihilation operator $\hat{a}_{A}$, frequency $\omega_{A}$),
a mechanical mode ($\hat{b}$, $\Omega_{0}$) and another optical
mode ($\hat{a}_{B}$, $\omega_{B}\neq\omega_{A}$). The mechanical
mode couples optomechanically to both optical modes. A and B are driven
by a laser with frequency $\omega_{L1}$ and $\omega_{L2}$, respectively:
For mode $A$, we require $\omega_{A}-\omega_{L1}=\Omega_{0}+\delta\equiv\Omega$
where $\omega_{L1}$ denotes the driving laser's frequency and $\delta\ll\Omega_{0}$
is the detuning from the red sideband. For mode $B$, a similar relation
$\omega_{B}-\omega_{L2}=\Omega$ holds, as depicted in the spectrum
in Figure \ref{fig:Schemes}(d). After application of the standard
linearization and RWA procedure \cite{Aspelmeyer2013RMPArxiv}, the
dynamics in a frame rotating with the drive is governed by the Hamiltonian
\begin{equation}
\hat{H}/\hbar=\Omega\sum_{i=A,B}\hat{a}_{i}^{\dagger}\hat{a}_{i}+\Omega_{0}\hat{b}^{\dagger}\hat{b}-(g_{A}^{*}\hat{a}_{A}\hat{b}^{\dagger}+g_{B}\hat{a}_{B}^{\dagger}\hat{b}+\text{h.c.}).\label{eq:HWavelengthConversion}
\end{equation}
Elimination of the mechanical mode leads to an effective Hamiltonian
$\hbar J_{\text{eff}}e^{i\phi}\hat{a}_{B}^{\dagger}\hat{a}_{A}+\text{h.c.}$
to leading order in $|g_{A,B}|/\delta$, with effective hopping rate
$J_{\text{eff}}=|g_{A}||g_{B}|/\delta$ and hopping phase \textbf{$\phi=\phi_{B}-\phi_{A}$
}. Here, $\phi_{A}$ and $\phi_{B}$ are the phases of the linearized
optomechanical interaction, of the form $g_{A}=|g_{A}|e^{i\phi_{A}}$,
which are set by the phase of the laser drive at the corresponding
site. Connecting alternating A and B sites by mechanical link modes
yields a row whose spectrum is depicted in Figure \ref{fig:Schemes}
(d). As in the previous scheme, we can simply connect rows by photonic
hopping without phases (at a rate $J_{\text{eff}}$) to yield a 2D
grid. Phase front engineering of the two driving lasers is sufficient
to realize arbitrary magnetic fields for photons in the grid. We note
that the scheme also works for driving far away from the red sideband
(yielding enhanced values of $\Omega$ and thereby $J_{{\rm eff}}$;
see below), though that requires stronger driving. 

Another optomechanical scheme for non-reciprocal photon transport
that could potentially be extended to a lattice is based on optical
microring resonators \cite{Hafezi2012OptExpr}, but the connection
of these rather large rings via waveguides would presumably result
in a more complicated and less compact structure than what can be
done with the photonic-crystal based approaches analyzed here. 

We now discuss the limitations imposed on the achievable effective
hopping $J_{{\rm eff}}$. The important end result will be that $J_{{\rm eff}}$
is limited to about the mechanical frequency $\Omega_{0}$, even though
perturbation theory would seem to imply a far smaller limit (for possible
technical limitations connected to the driving strength, see the Supplementary
Information).

We denote as $\epsilon\ll1$ the order of the three small parameters
$J/|\omega_{A}-\omega_{I}|$, $J/|\omega_{B}-\omega_{I}|$, and $g_{0}|\beta|/\Omega$
in the modulated link scheme (Fig. \ref{fig:Schemes} a,b). Then the
effective coupling strength in the perturbative regime reads $J_{\text{eff}}=\mathcal{O}(\epsilon^{3})\Omega$.
Even though the modulation frequency $\Omega$ need not equal the
eigenfrequency $\Omega_{0}$, they should usually be close to yield
a significant mechanical response and avoid other resonances. For
the wavelength conversion scheme, where $|g_{A,B}|/\delta=\mathcal{O}(\epsilon)$,
we recover $J_{\text{eff}}=\mathcal{O}(\epsilon^{2})\delta=\mathcal{O}(\epsilon^{3})\Omega_{0}$,
since RWA requires $\delta/\Omega_{0}$ to be small as well. In any
experimental realization, photons will decay at the rate $\kappa$.
Thus they travel $\sim J_{\text{eff}}/\kappa\sim\left(\mbox{\ensuremath{\Omega}}_{0}/\kappa\right)\mathcal{O}(\epsilon^{3})$
sites. In order for the photons to feel the magnetic field (or to
find nontrivial transport at all), this number should be larger than
1. That precludes being in the deep perturbative limit $\epsilon\ll1$,
even for a fairly well sideband-resolved system (where typically $\kappa\sim0.1\Omega_{0}$).
Similar considerations apply for other proposed (non-optomechanical)
schemes based on modulation \cite{Fang2012}. 

We now explore numerically the full dynamics, beyond the perturbative
limit. The optical local density of states (LDOS) is experimentally
accessible by measuring the reflection when probing an optical defect
mode via a tapered fiber, and it reveals the spectrum of the Hamiltonian.
It thus provides a reasonable way to asses the validity of the effective
Hamiltonian beyond the perturbative limit. Figure \ref{fig:OptomechanicalButterfly}
(a) shows the LDOS in the bulk calculated with the ideal effective
Hofstadter model (\ref{eq:HopEffective}) for a spatially constant
magnetic field, depicting the famous fractal Hofstadter butterfly
structure \cite{Hofstadter1976}. For comparison, we plot the LDOS
of the modulated link scheme in Figure \ref{fig:OptomechanicalButterfly}(b,c).
It is obtained by calculating numerically the Floquet Green's function
of the full equations of motion (with time-periodic coefficients),
see Appendix \ref{sec:FloquetGreen'sfunction}.  The results indicate
that the scheme works even for $J_{\text{eff}}\sim0.1\Omega>\kappa$,
although perturbation theory clearly breaks down in this regime. We
stress that the butterfly in Fig.~\ref{fig:OptomechanicalButterfly}(b,c)
could even be observed experimentally at room temperature, since the
spectrum is insensitive to thermal fluctuations. One would also observe
sidebands, see Figure \ref{fig:OptomechanicalButterfly}(d). Similar
results hold for the wavelength-conversion scheme (not shown here).

\begin{figure}[t]
\includegraphics[width=1\columnwidth]{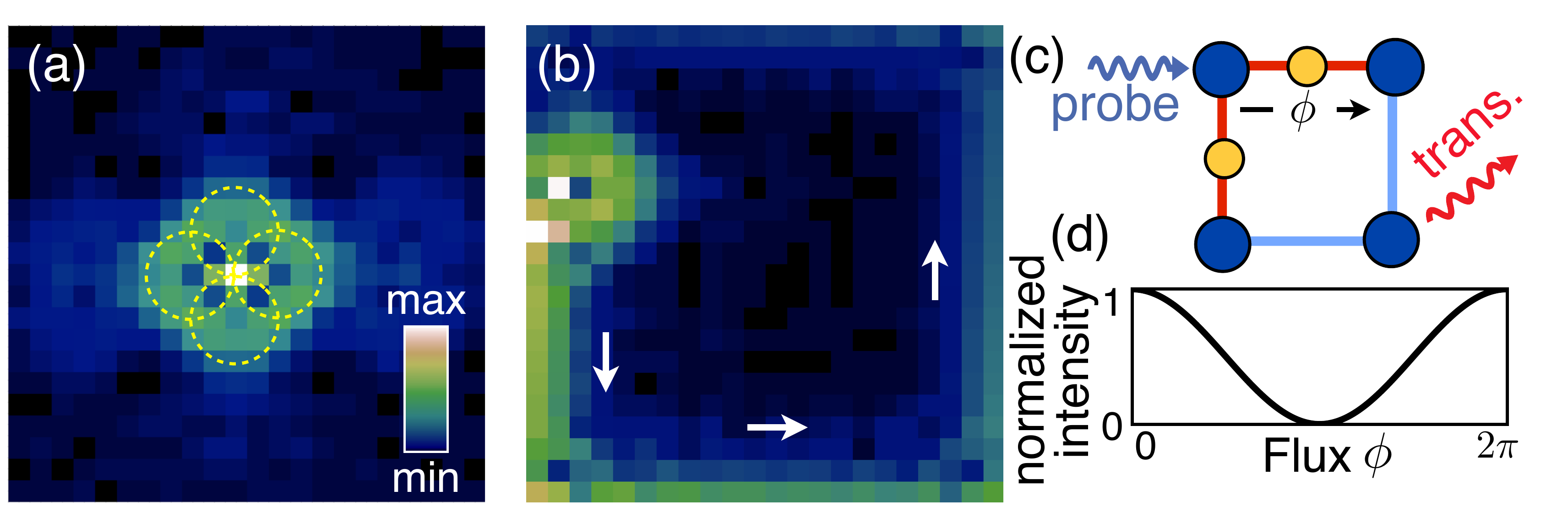}\protect\caption{Microscopic simulation of the wavelength conversion scheme, Eq. (\ref{eq:HWavelengthConversion}),
indicating its feasibility: Spatial distribution of light intensity
upon local injection of a probe laser in the bulk (a) and at the edge
(b), for a constant artificial magnetic field.  Bulk transport (a)
is governed by Landau levels and can be understood as a superposition
of classical cyclotron orbits (yellow circles) for different momentum
directions. (b) At the edges robust edge channels exist. (c) Optical
Aharonov-Bohm effect in minimal symmetric setup: (d) The interference
pattern (normalized probe laser transmission intensity) is shifted
by the magnetic flux through the ring. {[}Parameters: 22x22 grid (a,b);
$\delta=0.3\Omega_{0}$ (a,b), $\delta=0.1\Omega_{0}$ (d); $g=0.2\Omega_{0}$
(a,b), $g=0.01\Omega_{0}$ (d); $\kappa=0.01\Omega_{0}$; $\Gamma=\kappa/10$;
$\Phi=2\pi/8$ (a,b); $J=0.13\Omega_{0}$ (a,b), $J=0.001\Omega_{0}$
(d), $\Delta_{p}=1.278\Omega_{0}$ (a), $\Delta_{p}=1.260\Omega_{0}$
(b), $\Delta_{p}=1.103\Omega_{0}$ (d){]} \label{fig:TransportInUniformMagneticField}}
\end{figure}
In addition to measuring the optical spectrum, it is also possible
to look at photon transport in a spatially resolved manner, by injecting
a probe laser locally and then imaging the photons leaving the sample.
This provides another way to observe the effects of the artificial
gauge field, which gives rise to distinct transport phenomena as depicted
in Figure \ref{fig:TransportInUniformMagneticField}(a,b). For small
magnetic fields, $|\phi|\ll2\pi$, the dynamics can be understood
in the continuum limit when probing the bulk: One recovers the standard
Landau level picture for electrons in a constant magnetic field \cite{Hofstadter1976,LandauStatPhys2},
with effective mass $m^{*}=\hbar/2Ja^{2}$ and cyclotron frequency
$\omega_{\text{cyc}}=2\phi J$, where $a$ is the lattice constant.
In Figure \ref{fig:TransportInUniformMagneticField} (a)  the $n=1$
Landau level is selected via the probe's detuning $\Delta_{p}$ with
respect to the drive. The circles indicate the semi-classical cyclotron
orbits with radius $R=a\sqrt{(2n+1)/\phi}$. In this semi-classical
picture, the momentum of a photon injected locally at a site in the
bulk is equally distributed over all directions, since the position
is well-defined. Thus, the observed response resembles a superposition
of semi-classical circular Lorentz trajectories with different initial
velocity directions. A probe injected closer to the edge excites chiral
integer Quantum Hall Effect edge states, see Fig.~\ref{fig:TransportInUniformMagneticField}(b). 

The Aharonov-Bohm effect \cite{Aharonov1959} is one of the most intriguing
features of quantum mechanics. In an interferometer, electrons can
acquire a phase difference determined by the magnetic flux enclosed
by the interfering pathways, even though they never feel any force
due to the magnetic field. Figure \ref{fig:TransportInUniformMagneticField}(c)
depicts a setup that is based on the wavelength conversion scheme
and realizes an optical analog of the Aharonov-Bohm effect: A local
probe is transmitted via two pathways, leading to an interference
pattern in the transmission. The pattern is shifted according to the
flux through the 'ring', see Fig.~\ref{fig:TransportInUniformMagneticField}
(d), confirming the effect. 

All the effects displayed in Fig.~\ref{fig:TransportInUniformMagneticField}
have been simulated numerically for the wavelength conversion scheme,
see Appendix \ref{sec:detailsfrequency-conversion}, but similar results
hold for the modulated-link scheme.

\begin{figure}
\includegraphics[width=1\columnwidth]{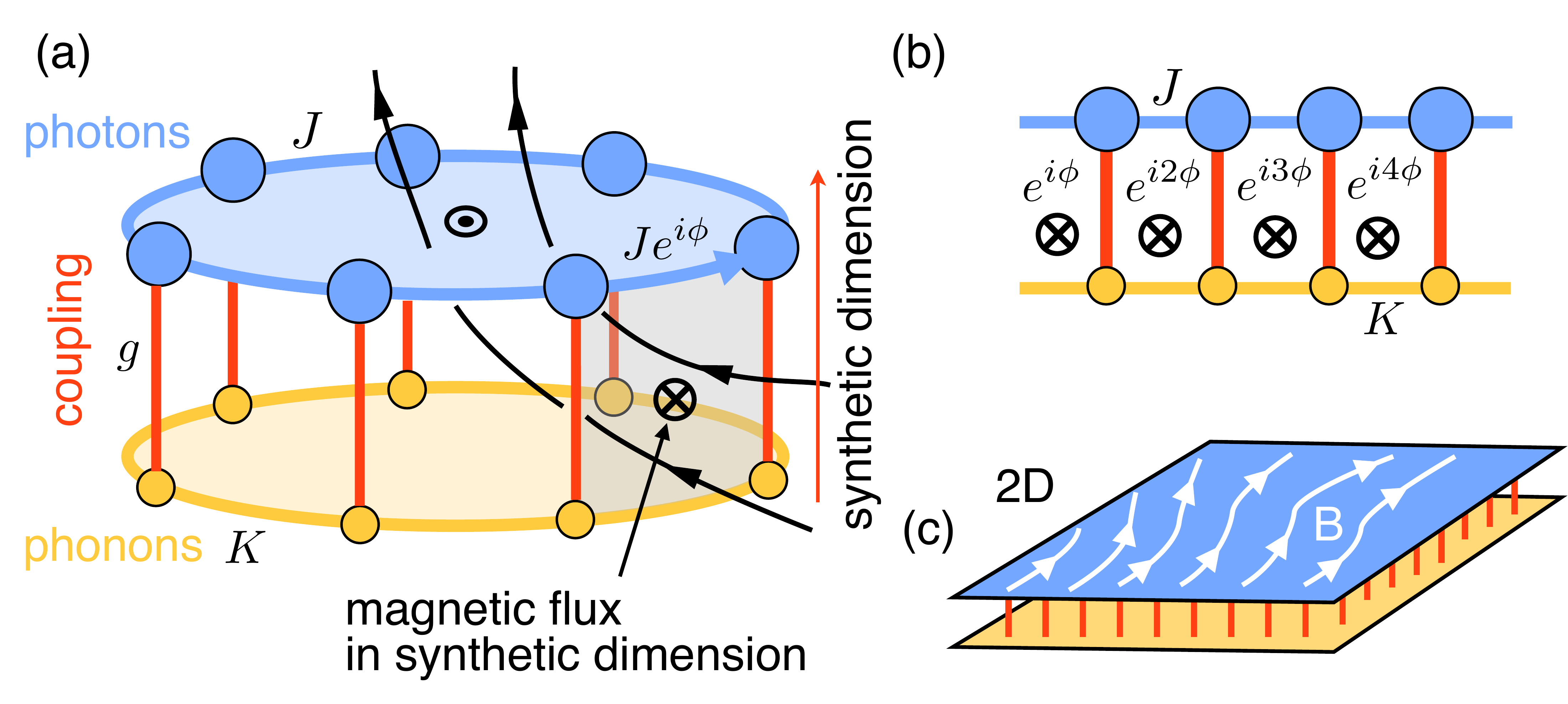}

\protect\caption{Optomechanical gauge fields within the concept of synthetic dimensions.
(a) The optomechanical coupling, $g$, can be viewed as connecting
sites along a synthetic dimension (photons vs. phonons). A phase for
the photon hopping, engineered using the schemes from above, creates
a flux in the optical plaquette (blue, top) \emph{and} in the adjacent
synthetic plaquette (gray). Hence, the magnetic field (black lines)
in the full space is divergence-free. (b) Engineering exclusively
the phases of $g$ allows to create magnetic fields/fluxes, but only
perpendicular to the synthetic dimension. Shining a single tilted
laser on a 1D chain yields a synthetic optomechanical ladder system
with constant synthetic magnetic flux. (c) 2D array, with the field
inside the (physical) plane generated by an arbitrary laser phase
pattern.\label{fig:GaugeFieldSyntheticDimension}}
\end{figure}
So far we have analyzed schemes to engineer hopping phases for photons.
We now ask about situations where the phonons are not only employed
as auxiliary virtual excitations, but rather occur as real excitations,
which can be interconverted with the photons. This means, in addition
to the modes making up the lattices described above (in either of
the two schemes), we now consider on-site vibrational modes $\hat{b}_{j}$
coupled optomechanically to the corresponding optical modes $\hat{a}_{j}$.
Using the standard approach \cite{Aspelmeyer2013RMPArxiv}, we arrive
at a linearized optomechanical interaction of the form $-g\hat{a}_{j}^{\dagger}\hat{b}_{j}+\text{h.c.}$.
Moreover, to be general (and generate nontrivial features connected
to the gauge field structure), we will assume the neighboring phonon
modes may also be coupled, as described by a tight-binding Hamiltonian
of the form $-K\sum_{\langle ij\rangle}\hat{b}_{j}^{\dagger}\hat{b}_{i}+\text{h.c.}$.

When discussing the effects of gauge fields in such a setting, the
system is best understood within the concept of 'synthetic' dimensions
\cite{Boada2012,Celi2014}. The optomechanical interaction can be
viewed in terms of an extension of the 1D or 2D lattice into such
an additional synthetic dimension. In our case, this dimension only
has two discrete locations, corresponding to photons vs. phonons.
In that picture, the optomechanical interaction, converting photons
to phonons, corresponds to a simple hopping between sites along the
additional direction. Figure \ref{fig:GaugeFieldSyntheticDimension}
(a) sketches this for an optomechanical ring: photons and phonons
represent two layers separated along the synthetic dimension. Applying
any of our two previously discussed schemes, a photon hopping from
site $i$ to $j$ will acquire a phase $\phi_{ij}=\int_{\vec{r}_{i}}^{\vec{r}_{j}}d\vec{r}\vec{A}$.
The gauge field ${\bf A}$ must now be viewed as a vector field in
this new 3d space, where one of the dimensions is synthetic. A finite
hopping phase $\phi$ at one of the optical links creates a magnetic
flux through the optical plaquette as desired, see Fig.~ \ref{fig:GaugeFieldSyntheticDimension}
(a). However, and this is the important point, since the magnetic
field ${\bf B}$ is divergence-free, the field must penetrate at least
one additional plaquette, causing the opposite magnetic flux in the
synthetic dimension (assuming $g\in\mathbb{R}$). In general, realizing
that there is this kind of behaviour is crucial to avoid puzzles about
seeming violations of gauge symmetry in situations with photon magnetic
fields in optomechanical arrays. It is necessary to keep track of
the full vector potential in the space that includes the synthetic
dimension.

We now take a step back, getting rid of the previously discussed engineered
schemes that required two lasers and some arrangement of 'link' modes.
Rather we will consider simple optomechanical arrays, i.e. lattices
of optical and vibrational modes, with photon and phonon tunnel coupling
between modes and with the optomechanical interaction.  We ask: What
is the effect of an arbitrary, spatially varying optical phase field
in the driving laser that sets the strength of the optomechanical
coupling? It turns out that the resulting spatially varying phase
of the optomechanical coupling, $g_{j}=|g_{j}|e^{i\varphi_{j}}$,
can be chosen to create arbitrary magnetic fields perpendicular to
the synthetic dimension. A particularly simple example is a simple
linear chain of optomechanical cells. Shining a tilted laser (i.e.
with a phase gradient, $\varphi_{j}=j\cdot\delta\varphi$) onto such
a 1D optomechanical array creates a constant magnetic flux through
the  plaquettes of the ``optomechanical synthetic ladder'' that
can be drawn to understand the situation, cf. Figure \ref{fig:GaugeFieldSyntheticDimension}
(b). The quantum mechanics of excitations tunneling between the two
'rails' of the ladder (corresponding to photon-phonon conversion)
is directly analogous to experiments on electron tunneling between
parallel wires in a magnetic field \cite{Steinberg2008}. The magnetic
field shifts the momenta of the tunneling particles, giving rise to
resonance phenomena when the shifted dispersion curves $\omega(k)$
of the excitations match. Via phase front engineering one could create
arbitrary synthetic magnetic fields also in 2D grids, see Figure \ref{fig:GaugeFieldSyntheticDimension}(c).
We note, though, that this method is constrained since it cannot create
directly magnetic fluxes through optical or mechanical plaquettes,
and in general only the schemes discussed above provide full flexibility.
On the other hand, if either the photon or phonon modes are occupied
only virtually, then effective fluxes can still be generated for the
remaining real excitations, even with a single laser, and this works
best for phonons (see \cite{Habraken2012,Peano2014}).

We now discuss the most salient aspects of the experimental realization.
Both the 'butterfly' optical spectrum and spatially resolved transport
can be probed using homodyne techniques, which are insensitive to
noise. Real-space optical imaging is feasible, as the defects are
a few micrometers apart. The optical phase pattern can be engineered
using spatial light modulators. No time-dependent changes of the pattern
are needed here, since the time-dependence is generated via the beat-note
between the two laser beams. 

For the modulated-link scheme, the mechanical oscillation amplitude
$\beta$ used for the modulation should overwhelm any thermal fluctuations.
In the example of Fig.~\ref{fig:OptomechanicalButterfly}, we assumed
$g_{0}\beta=0.3\Omega_{0}$. At recently achieved parameters \cite{SafaviNaeini2014SnowCavity}
$g_{0}/2\pi=220\,\text{kHz}$ and $\Omega_{0}/2\pi=9\,\text{GHz}$
, this would imply $\beta\sim10^{4}$, i.e. a phonon number of $10^{8}$
reached by driving, certainly larger than the thermal population.
If we drive the mechanical vibration using a radiation pressure force
oscillating at resonance (assuming the quoted $g_{0}$ also for the
optical mode used in that driving), then we have $\beta=2g_{0}n_{c}/\Gamma$,
where $n_{c}$ is the circulating photon number and $\Gamma$ the
mechanical damping rate. Given a mechanical quality factor of $\Omega_{0}/\Gamma=2\cdot10^{5}$,
this requires $n_{c}\sim10^{3}$ photons for Fig.~\ref{fig:OptomechanicalButterfly},
a realistic number. We note that thermal fluctuations of the mechanical
amplitude give rise to a fractional deviation of $\sqrt{\bar{n}_{{\rm th}}}/\beta$
in $J_{{\rm eff}}$, with a slow drift on the time scale $\Gamma^{-1}$.
At typical temperatures used in experiments, we have $\bar{n}_{{\rm th}}\sim100-1000$,
and so the fractional change is on the order of a percent, which will
not noticeably impact transport.

In the alternative wavelength-conversion scheme, one should strive
for a large photon-enhanced optomechanical coupling rate $g=g_{0}\alpha$.
A general estimate implies we always need the photon number to be
larger than $(\kappa/g_{0})^{2}$ in order to see the butterfly spectrum
and the transport effects. This condition (compatible with Fig.~\ref{fig:TransportInUniformMagneticField})
would require a circulating photon number of around $3\cdot10^{5}$
for the parameters demonstrated in a recent successful wavelength
conversion setup based on optomechanical crystals \cite{Hill2012WavelengthConversion}.
It is also important to estimate the unwanted influx of thermal excitations
from the phonon subsystem into the photon subsystem, at least if the
setup is to be applied in the quantum regime, for observing the transport
of \emph{single} photons in the presence of a magnetic field. In the
wavelength-conversion scheme, there is a detuning $\delta\gg\kappa$
between the red sideband of the laser and the phonon mode, such that
photon-phonon conversion is suppressed. Nevertheless, it still happens
at a rate $\gamma\bar{n}$, where $\gamma=g^{2}/\delta$ is the ``cooling
rate'' (for the detuned case applicable here) and $\bar{n}$ is the
number of phonons in the mode. Fortunately, this phonon number is
also reduced by the very same off-resonant cooling process. Balancing
the inflow and outflow of excitations, we find that there will be
a remaining unwanted photon occupation of $\bar{n}_{{\rm phot}}^{{\rm th}}=\bar{n}_{{\rm th}}\Gamma/\kappa$
due to the conversion of thermal phonons into photons, where $\bar{n}_{{\rm th}}$
is the bulk thermal phonon occupation. The factor $\Gamma/\kappa$
suppresses this number strongly, and it should be possible to reach
the regime $\bar{n}_{{\rm phot}}^{{\rm th}}\ll1$ in low-temperature
setups.

Reducing fabrication-induced disorder will be crucial for any future
applications of photonic crystals, including the one envisaged here
(as well as other photonic magnetic field schemes). In first experimental
attempts, the optical and mechanical disorder is on the percent level,
which makes especially the fluctuations of the optical resonance frequencies
significant. Nevertheless, strong reductions of the disorder will
be possible by post-fabrication methods \cite{Zheng2011,Schmidt2009,Imamoglu2006},
such as local laser-induced oxidation. These are expected to reduce
the fluctuations down to the level of $10^{-5}$ relative optical
frequency fluctuations. This is enough to suppress the optical disorder
to some fraction of the photon hopping rate $J_{{\rm eff}}\sim\Omega$,
which will enable near-ideal photon transport (e.g. Anderson localization
lengths would be at least hundreds of sites, larger than the typical
arrays). Disorder in the mechanical frequencies can be reduced by
similar techniques, but is much less problematic, due to the difference
in absolute frequency scales between optics and mechanics.

\emph{Outlook} - Optomechanical crystals represent an interesting
system for the realization of artificial photonic magnetic fields
due to their all-optical controllability. Moreover, their rich non-linear
(quantum) dynamics \cite{Ludwig2013} could be explored in the presence
of an artificial magnetic field. In general, the very flexible optical
control could be used to create and explore novel features, e.g. varying
the optomechanical coupling strength spatially and/or temporally,
both adiabatically and with sudden quenches. Moreover, a second strong
control laser could be used to create a spatially and temporarily
varying optical on-site potential landscape. \textbf{\uline{}}

\emph{Acknowledgements} - This work was supported via an ERC Starting
Grant OPTOMECH, the ITN cQOM, the DARPA program ORCHID, and via the
Institute for Quantum Information and Matter, an NSF Physics Frontiers
Center with support of the Gordon and Betty Moore Foundation.

\appendix

\section{Derivation of the effective magnetic Hamiltonian for the modulated
link scheme\label{sec:detailsFloquetHamiltonian}}

Here, we derive the effective Hamiltonian $\hbar J_{{\rm eff}}e^{i\phi}\hat{a}{}_{B}^{\dagger}\hat{a}{}_{A}+\text{h.c.}$
that describes the tunneling of photons from site A to site B in the
presence of an effective magnetic field created using the modulated
link scheme. We start from the full time dependent Hamiltonian Eq.~(\ref{eq:FloquetSchemeTimeDepH}).
Since this second-quantized Hamiltonian is particle conserving we
can switch to a first-quantized picture in the standard way. The corresponding
single-particle Hamiltonian $\hat{H}_{M}$ reads 
\[
\hat{H}_{M}=\hbar\begin{pmatrix}\omega_{A} & 0 & -J\\
0 & \omega_{B} & -J\\
-J & -J & \bar{\omega}_{I}+2g_{0}|\beta|\cos(\Omega t+\phi)
\end{pmatrix}.
\]
It acts on the photon wavefunction $|\psi\rangle\equiv(\psi_{A},\psi_{B},\psi_{I})$
where $\psi_{s}$ describes the probability amplitude that the photon
is localized on site $s$, $s=A,B,I$. Since the Hamiltonian is time
periodic, there is a complete set of quasi-periodic solutions of the
Schr�dinger equation, $|\psi_{j}(t+2\pi/\Omega)\rangle=\exp[-i2\pi\omega_{j}/\Omega]|\psi_{j}(t)\rangle$
where $2\pi/\Omega$ is the period and index $j$ spans the Hilbert
space, $j=1,2,3$. In practice, one solves the eigenvalue problem
$\varepsilon_{jm}\big|\phi_{jm}\rangle\rangle={\cal H}{}_{jm}\big|\phi_{jm}\rangle\rangle$
where ${\cal H}\equiv-i\hbar\partial_{t}+\hat{H}_{M}$ is the Floquet-Hamiltonian,
$\varepsilon_{jm}=\hbar(\omega_{j}+m\Omega)$ are the quasienergies
and $\big|\phi_{jm}\rangle\rangle=\exp[i(\omega_{j}+m\Omega)t]\big|\psi_{j}(t)\rangle$
are time-periodic states, the so-called Floquet eigenstates {[}$m\in\mathbb{Z}${]}
\cite{SAMBE1973}. Notice that the Floquet Hamiltonian can be regarded
as an operator on the extended Hilbert space of the time periodic
vectors equipped with the scalar product 
\begin{equation}
\langle\langle\phi_{j}\big|\phi_{m}\rangle\rangle=\frac{1}{T}\int_{0}^{T}\langle\phi_{j}(t)|\phi_{m}(t)\rangle.\label{eq:Floquetscalarproduct}
\end{equation}

\begin{figure}
\includegraphics[width=1\columnwidth]{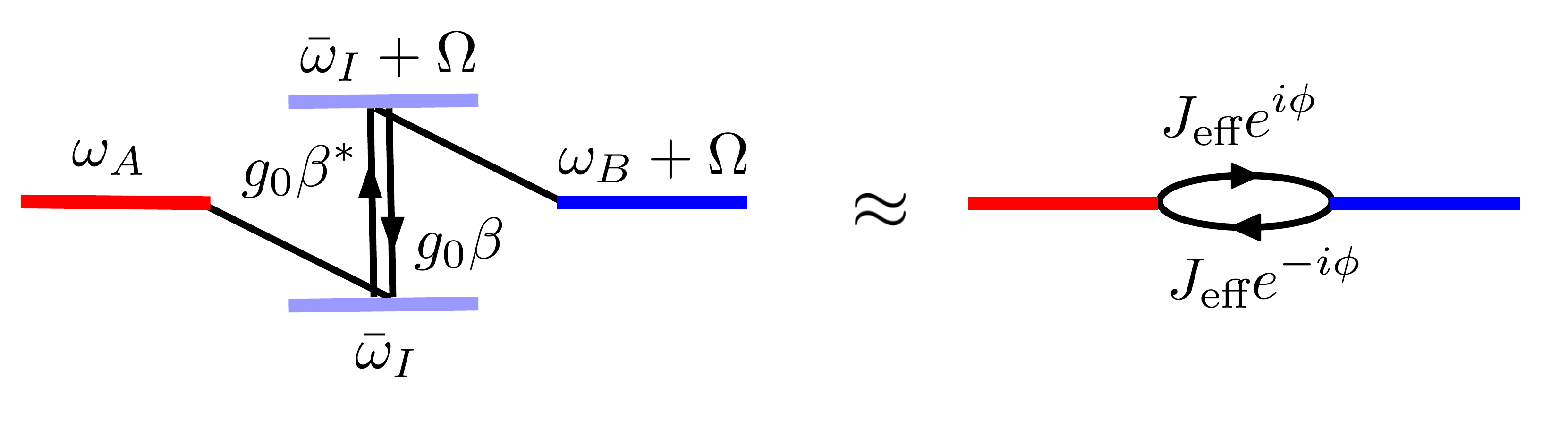}\protect\caption{\label{fig:Floquet-level-scheme-INPAPER-2}Floquet level scheme, i.e.
quasienergy levels, describing the hopping transitions between sites
$A$ and $B$ mediated by the virtual transition trough site $I$
accompanied by the exchange of a phonon.}
\end{figure}

In this framework, we can use the standard quantum mechanical perturbation
theory to derive an effective time independent single-particle Hamiltonian.
We assume a resonant drive $\omega_{A}\approx\omega_{B}+\Omega$,
and weak tunneling/driving, $J_{A,B}/|\omega_{A,B}-\bar{\omega}_{I}|$,
$g_{0}\beta/\Omega\ll1$. We identify resonant Floquet-levels with
quasienergies $\hbar\omega_{A}$ and $\hbar(\omega_{B}+\Omega$) coupled
via the third order virtual tunneling process through the interface
site I shown in Figure \ref{fig:Floquet-level-scheme-INPAPER-2}.
Up to leading order in perturbation theory, we can focus on the block
of the Floquet Hamiltonian comprising the four unperturbed quasienergy
levels that are involved in this process, cf. Figure \ref{fig:Floquet-level-scheme-INPAPER-2},
\[
{\cal \hat{H}}=\hbar\begin{pmatrix}\omega_{A} & 0 & -J & 0\\
0 & \omega_{B}+\Omega & 0 & -J\\
-J & 0 & \omega_{I} & g_{0}\beta\\
0 & -J & g_{0}\beta^{*} & \omega_{I}+\Omega
\end{pmatrix}.
\]
Application of a standard Schrieffer-Wolff transformation \cite{Schrieffer1966,Shavitt1980,Bravyi2011},
i.e. applying degenerate perturbation theory to third order, leads
to the effective block diagonal Floquet Hamiltonian
\begin{equation}
{\cal \hat{H}}_{{\rm eff}}=\hbar\begin{pmatrix}\tilde{\omega}_{A} & J_{{\rm eff}}e^{-i\phi}\\
J_{{\rm eff}}e^{i\phi} & \tilde{\omega}_{B}+\omega_{{\rm ex}}
\end{pmatrix}.\label{eq:hameffectivesingle-particle-1}
\end{equation}
where $\tilde{\omega}_{s}=\omega_{s}+J_{s}^{2}/(\omega_{s}-\bar{\omega}_{I})$
with $s=A,B$ and $J_{{\rm eff}}=g_{0}|\beta|J_{A}J_{B}/[(\omega_{A}-\bar{\omega}_{I})(\omega_{B}-\bar{\omega}_{I})]$.
Finally, we turn back Hamiltonian \ref{eq:hameffectivesingle-particle-1}
into its second-quantized form and switch to a frame rotating with
frequency $\tilde{\omega}_{A}$ ($\tilde{\omega}_{B}$) on site A
(B). For a resonant drive, $\tilde{\omega}_{A}=\tilde{\omega}_{B}+\Omega$,
this yields the desired form of the second-quantized effective Hamiltonian,
$J_{{\rm eff}}(e^{-i\phi}\hat{a}{}_{A}^{\dagger}\hat{a}{}_{B}+e^{i\phi}\hat{a}{}_{B}^{\dagger}\hat{a}{}_{A}).$

\section{Transmission amplitudes and density of states for the modulated link
scheme\label{sec:FloquetGreen'sfunction}}

Here, we calculate the LDOS for the modulated link scheme which is
plotted in Figure \ref{fig:OptomechanicalButterfly}. We use the full
time dependent Hamiltonian Eq.~(\ref{eq:FloquetSchemeTimeDepH})
extended to the whole lattice (including also the sublattice formed
by the link sites). Since we are dealing with a time periodic system
where the energy is not a constant of motion, we have to appropriately
generalize the definition of the LDOS. A natural generalization of
the standard definition to time-periodic systems is the following,
\[
\rho(\omega,\mathbf{j})=-2{\rm Im}G(\omega,0;\mathbf{j},\mathbf{j})
\]
where $G(\omega,m;\mathbf{j},\mathbf{l})$ is the Floquet Green's
function
\[
G=\frac{-i}{T}\int_{0}^{T}\!\! d\tau\int_{0}^{\infty}\! dte^{i(\omega+m\Omega)t+im\Omega\tau}\langle[\hat{a}_{\mathbf{j}}(t+\tau),\hat{a}_{\mathbf{l}}^{\dagger}(\tau)]\rangle.
\]
The Floquet Green's function describes the (linear) response of the
array to a probe laser. More precisely, the light amplitude on site
$\mathbf{j}$ in the presence of a probe drive on site $\mathbf{l}$
with frequency $\omega$ and amplitude $\alpha^{(in)}$ {[}described
by the additional Hamiltonian term $H_{I}=i\hbar\sqrt{\kappa}\alpha^{(in)}(\hat{a}_{\mathbf{l}}^{\dagger}e^{-i\omega t}+h.c.)${]}
 is 
\[
\langle\hat{a}_{\mathbf{j}}(t)\rangle=\sum_{m}e^{-i(\omega+m\Omega)t}i\sqrt{\kappa}\alpha^{(in)}G(\omega,m;\mathbf{j},\mathbf{l}).
\]
This is essentially a generalization of the Kubo formula which applies
to any time periodic Hamiltonian. Using the input-output relations,
$\hat{a}_{\mathbf{j}}^{(out)}(t)=\hat{a}_{\mathbf{j}}^{(in)}(t)-\sqrt{\kappa}\hat{a}_{\mathbf{j}}$,
we can also calculate the field outside the cavity,
\[
\langle\hat{a}_{\mathbf{j}}^{(out)}(t)\rangle\equiv\sum_{m}e^{-i(\omega+m\Omega)t}t_{O}(\omega,m;\mathbf{j},\mathbf{l})\alpha^{(in)},
\]
where
\[
t_{O}(\omega,m;\mathbf{j},\mathbf{l})=\delta_{\mathbf{j\mathbf{l}}}\delta_{m0}-i\kappa G(\omega,m;\mathbf{j},\mathbf{l})
\]
is the transmission amplitude of a photon from site $\mathbf{l}$
to site $\mathbf{j}$ if it has been up-converted $m$-times (or down-converted
$|m|$-times for $m$ negative). 

For a time-periodic system with a particle conserving Hamiltonian,
the Floquet Green's function can be easily expressed in terms of the
first-quantized Floquet Hamiltonian ${\cal H}=-i\partial_{t}-H(t)$,
\[
G(\omega,m;\mathbf{j},\mathbf{l})=\big\langle\big\langle\mathbf{j},m\big|\left(\omega-{\cal H}+i\kappa/2\right)^{-1}\big|\mathbf{l},0\big\rangle\big\rangle
\]
Notice that the Floquet Hamiltonian and the Green's function can be
regarded as operators acting on the extended Hilbert space of the
time-periodic photon states with the scalar product Eq.~(\ref{eq:Floquetscalarproduct}
). As such they acts on the time periodic states $|\mathbf{j},m\rangle\rangle,$
where index $\mathbf{j}$ indicates the lattice site and $m$ the
Fourier component. Thus, the density of states can be readily computed
by diagonalizing the Floquet Green's function. We find 
\[
\text{\ensuremath{\rho}}(\omega)=\sum_{k}\frac{\kappa}{(\omega-\omega_{k})^{2}+\kappa^{2}/4}\Big|\big\langle\big\langle\mathbf{j},0\big|\phi_{k}\big\rangle\big\rangle\Big|^{2}\,,
\]
where $\hbar\omega_{k}$ are the quasienergies and $\big|\phi_{k}\rangle\rangle$
are the corresponding Floquet eigenstates obtained by numerically
diagonalizing ${\cal H}$. Taking into account that the Floquet eigenfunctions
$\big|\phi_{k}\rangle\rangle$ forms a complete orthonormal basis
of the Hilbert space of the time-periodic states {[}with the scalar
product Eq.~(\ref{eq:Floquetscalarproduct} ){]}, it immediately
follows that the density of states is appropriately normalized,
\[
\int_{-\infty}^{\infty}d\omega\rho(\omega)=2\pi.
\]

\section{Transmission amplitudes for the frequency-conversion\emph{ }scheme\label{sec:detailsfrequency-conversion}}

For the frequency-conversion scheme we start from the linearized Langevin
equations for the full array including the mechanical links modes
\cite{Aspelmeyer2013RMPArxiv,Schmidt2014},
\begin{eqnarray}
\dot{\hat{b}}_{\mathbf{k}} & = & i\hbar^{-1}[\hat{H},\hat{b}_{\mathbf{k}}]-\Gamma\hat{b}_{\mathbf{k}}/2+\sqrt{\Gamma}\hat{b}_{\mathbf{k}}^{({\rm in)}},\nonumber \\
\dot{\hat{a}}_{\mathbf{j}} & = & i\hbar^{-1}[\hat{H},\hat{a}_{\mathbf{j}}]-\kappa\hat{a}_{\mathbf{j}}/2+\sqrt{\kappa}\hat{a}_{\mathbf{j}}^{({\rm in)}}.\label{eq:Langevinfreqconv}
\end{eqnarray}
The first line (second line) describes the sites hosting a mechanical
(optical) mode. The Hamiltonian $\hat{H}$ is given by Eq. (\ref{eq:HWavelengthConversion})
extended to the full array and the noise forces have the usual commutation
relations \cite{Aspelmeyer2013RMPArxiv}. Notice that Eq. (\ref{eq:Langevinfreqconv})
is written in a frame where the optical modes on sublattice A and
B are rotating with frequency $\omega_{L1}$ and $\omega_{L2}$, respectively.
A probe laser on site $\mathbf{l}$ with frequency $\omega$ and amplitude
$\alpha^{in}$ is described by the additional Hamiltonian term $H_{I}=i\hbar\sqrt{\kappa}\alpha^{(in)}(\hat{a}_{\mathbf{l}}^{\dagger}e^{-i\Delta_{p}t}-h.c.)$,
where $\Delta_{p}=\omega-\omega_{Ls}$ ($s=1,2$ for $\mathbf{l}$
on sublattice A or B, respectively). The linear response of the light
amplitude on site $\mathbf{j}$ to such probe laser is given by the
Kubo formula
\begin{eqnarray}
\langle\hat{a}_{\mathbf{j}}(t)\rangle & = & i\sqrt{\kappa}\alpha^{(in)}e^{-i\Delta_{p}t}G_{\hat{a}\hat{a}^{\dagger}}(\Delta_{p},\mathbf{j},\mathbf{l})\nonumber \\
 &  & -i\sqrt{\kappa}\alpha^{(in)}e^{i\Delta_{p}t}G_{\hat{a}\hat{a}}(-\Delta_{p},\mathbf{j},\mathbf{l}),\label{eq:Kuboformula}
\end{eqnarray}
with the Green's functions
\begin{eqnarray*}
G_{\hat{a}\hat{a}^{\dagger}}(\omega,\mathbf{j},\mathbf{l}) & =-i & \int_{0}^{\infty}\! dte^{i\omega t}\langle[\hat{a}_{\mathbf{j}}(t),\hat{a}_{\mathbf{l}}^{\dagger}(0)]\rangle,\\
G_{\hat{a}\hat{a}}(\omega,\mathbf{j},\mathbf{l}) & =-i & \int_{0}^{\infty}\! dte^{i\omega t}\langle[\hat{a}_{\mathbf{j}}(t),\hat{a}_{\mathbf{l}}(0)]\rangle.
\end{eqnarray*}
Notice that in Figure 3 and 4 of the main text we plot the resonant
part of the response corresponding to the first line of Eq. (\ref{eq:Kuboformula}).
If $\mathbf{j}$ and $\mathbf{l}$ lie on different sublattices, the
frequency of the probe signal is converted {[}to read off this frequency
from Eq. (\ref{eq:Kuboformula}), one has to keep in mind that the
frame of reference is rotating at different frequencies on the two
optical sublattices{]}. Finally, we note that the light transmitted
outside of the sample $\left\langle \hat{a}_{\mathbf{j}}^{({\rm out})}\right\rangle =t(\omega,\mathbf{j},\mathbf{l})\left\langle \hat{a}_{\mathbf{l}}^{({\rm in})}\right\rangle $
can be readily computed using the input output relations \cite{WallsMilburn_QuantumOptics}
$\hat{a}_{\mathbf{j}}^{({\rm out})}=\hat{a}_{\mathbf{j}}^{({\rm in})}-\sqrt{\kappa}\hat{a}_{\mathbf{j}}$.
From Eq. (\ref{eq:Kuboformula}) we find the transmission amplitude
\begin{equation}
t_{O}(\omega,\mathbf{l},\mathbf{j})=\delta_{\mathbf{lj}}-i\kappa G_{\hat{a}\hat{a}^{\dagger}}(\omega,\mathbf{l},\mathbf{j}).\label{eq:transmission}
\end{equation}
Since the transmission amplitudes of a probe laser beam are \emph{generally}
proportional to the corresponding light amplitudes inside the array
(on all sites except for the one where the light is injected), the
amplitude patterns shown in Figures 3 and 4 could be directly measured
by a position resolved measurement of the light scattered by the array.

In order to calculate the transmission in Figures 3 and 4 we have
calculated the Green's function numerically. We note that for an array
with $N\times N$ optical sites, there is a total of $N(2N-1)$ sites
(including also the mechanical sites) and a total of $2N(2N-1)$ degrees
of freedom. Thus, computing numerically the Green's function amounts
to inverting a $2N(2N-1)\times2N(2N-1)$ matrix. In Figure 3 and 4
we have chosen $N=22$.

\end{document}